# Quantifying the local mechanical properties of twisted double bilayer graphene

Alessandra Canetta,[†] Sergio Gonzalez-Munoz,[‡] Khushboo Agarwal,[‡] Pauline de Crombrugghe,[†] Viet-Hung Nguyen,[¶] Yuanzhuo Hong,[§] Sambit Mohapatra ,[§] Kenji Watanabe,[∥] Takashi Taniguchi,[∥] Bernard Nysten,[†] Benoît Hackens,[†] Rebeca Ribeiro-Palau,[§] Jean-Christophe Charlier,[¶] Oleg Kolosov,[*,‡] Jean Spièce,[*,†] and Pascal Gehring[*,†]

[†]*IMCN/NAPS, Université Catholique de Louvain (UCLouvain), 1348 Louvain-la-Neuve, Belgium*
[‡]*Physics Department, Lancaster University, Lancaster, UK*
[¶]*IMCN/MODL, Université Catholique de Louvain (UCLouvain), 1348 Louvain-la-Neuve, Belgium*
[§]*Centre de Nanosciences et de Nanotechnologies (C2N), CNRS, Université Paris Sud, Université Paris-Saclay, Palaiseau, France*
[∥]*National Institute for Materials Science, 1-1 Namiki, Tsukuba 305-0044, Japan*

E-mail: o.kolosov@lancaster.ac.uk; jean.spiece@uclouvain.be; pascal.gehring@uclouvain.be

## Abstract

Nanomechanical measurements of minimally twisted van der Waals materials remained elusive despite their fundamental importance for device realisation. Here, we use Ultrasonic Force Microscopy (UFM) to locally quantify the variation of out-of-plane Young's modulus in minimally twisted double bilayer graphene (TDBG). We reveal a softening of the Young's modulus by 7% and 17% along single and double domain walls, respectively. Our experimental results are confirmed by force-field relaxation models. This study highlights the strong tunability of nanomechanical properties in engineered twisted materials, and paves the way for future applications of designer 2D nanomechanical systems.

## Introduction

Recent studies showed that the twist angle between van der Waals-stacked two-dimensional (2D) atomic layers strongly impacts their electrical, optical or magnetic properties.[1–7] Less investigated, however, is the influence of twisting on nanomechanical properties.[8–10] Strain accumulation at domain boundaries in the moiré superlattice is expected to have significant impact on the mechanical properties of twisted materials,[11] which could play a role in the frustration of flat band formation[11,12] or the formation of a stacking domain structure.[13] However, while the commensurate-incommensurate transition in graphene on hexagonal boron nitride has been reported using nanomechanical microscopy,[14] a full quantification of nanomechanics in twisted 2D materials is to date missing.

Here, we use the Ultrasonic Force Microscopy (UFM)[15,16] method (see Figure 1), which is based on contact Atomic Force Microscopy (AFM), to investigate quantitatively the local mechanical properties of minimally Twisted Double Bilayer Graphene (TDBG). TDBG provides a rich testbed for local mechanical studies as it contains single and double domain boundaries separating commensurate stacking domains.[13] This allows us to reveal local variation of out-of-plane Young's modulus induced by stacking-domain boundaries of the moiré pattern. Our results are compared to force-field relaxation models, which compute the structural morphology of the twisted pattern and Young's modulus variations across the sample.

Ultrasonic Force Microscopy (UFM) directly probes the material stiffness under the AFM tip thanks to a high frequency mechanical actuation of the sample and non-linear detection of ultrasonic vibrations.[17] In the UFM setup, a ceramic piezoelectric transducer is used to mechanically vibrate the sample at ultrasonic frequency ($\sim$ 4 MHz) (see Figure 1a). At such frequency, much higher than the cantilever fundamental free resonance, the probe cannot follow the sample vibration. Thus, the cantilever becomes effectively stiff with a spring constant around $10^4$ Nm$^{-1}$.[18] As a consequence, the tip is able to indent into both soft and hard materials.[18–20]

Figure 1b shows the dependence of the normal force $F$ from the indentation $h$. The latter is modulated sinusoidally around $h_1$. For small oscillations ($\Delta h_0$), the normal force averaged over one modulation cycle remains equal to $F_1$ (see Fig. 1b and c) and no detectable cantilever displacement is produced. Nevertheless, when the excitation amplitude is increased to $\Delta h_1$, the contact breaks and the average force is changed.[18] The UFM signal is thus produced by modulating the ultrasonic excitation envelope with a diamond shaped signal (with a frequency of 2.7 kHz) which can be detected via the laser deflection on the photodetector, as shown on Fig.1a. As a result, the UFM response depends on the local materials properties thanks to the non-linear nature of the tip-sample contact.[16,18,21]

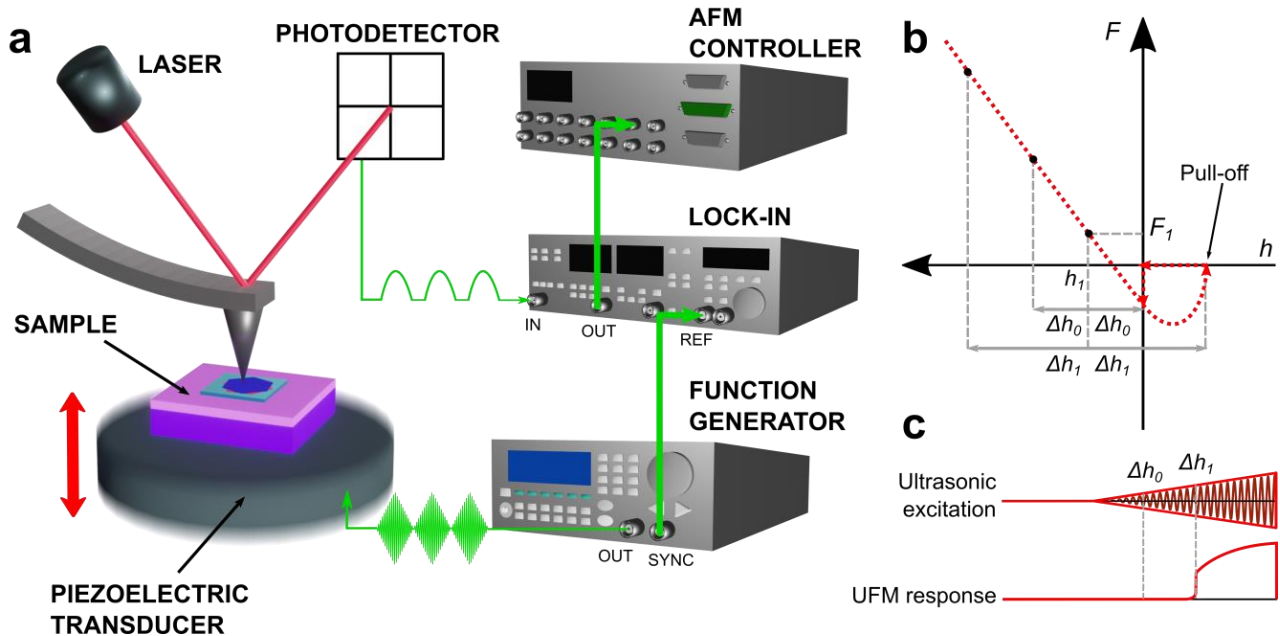

**Fig. 1. UFM method.** a) Schematic of the UFM setup. An AC-voltage in the MHz range modulated by a diamond shaped signal in the kHz range is applied to the piezoelectric transducer using a function generator. The red arrow shows the out-of-plane direction of the piezoelectric transducer vibration. The laser deflection from the cantilever is recorded by a photodetector, whose output signal is demodulated with a lock-in pre-amplifier. This UFM response is recorded by an AFM controller. b) Force-indentation curve. For small ultrasonic amplitudes ($\Delta h_0$), the normal force averaged in time over one ultrasonic period, at an indentation depth $h_1$, is equal to the initial value $F_1$ as the force curve can be considered linear. $\Delta h_1$ is the threshold amplitude necessary to reach the pull-off point. c) Schematic of the normal deflection response induced by the out-of-plane ultrasonic vibration of the sample. A change in normal deflection occurs only at ultrasonic amplitudes higher than $\Delta h_1$. The discontinuity occurring at this value is defined as force jump.[18,21]

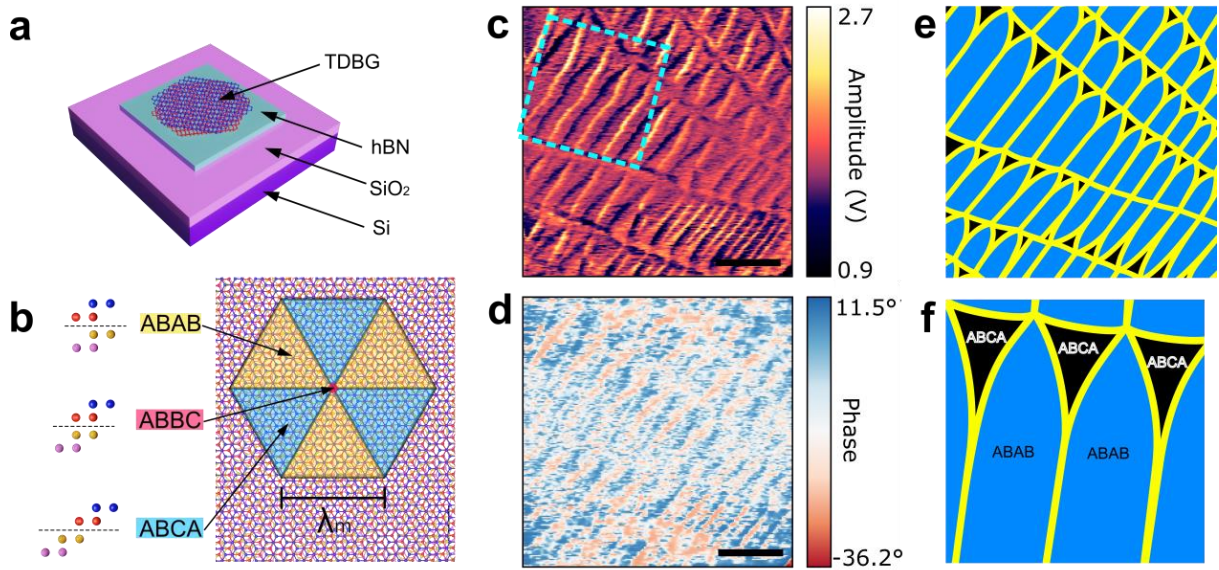

**Fig. 2. Stacking order domains in twisted double bilayer graphene and imaging of the relative moiré superlattice by PFM.** a) Schematic of the device, consisting in two AB-stacked bilayer graphene flakes (TDBG) overlapped with a twist angle, on top of a hexagonal boron nitride (hBN) flake on a Si/SiO₂ substrate. b) Side views of the main stacking configurations - ABAB, ABBC and ABCA respectively - in a TDBG moiré superlattice rotated by an angle $\vartheta$ and their relative locations in the pattern. $\lambda_m$ is the lattice constant of the moiré superlattice. c-d) PFM images (amplitude and phase, respectively) of TDBG. The maps show evidence of the moiré superlattice, in which different stacking domains can be distinguished. The squared region delimited by the cyan dashed line refers to the area where UFM measurements have been performed. Scale bars: 200 nm. e) Schematic of the stacking domain distribution of the pattern shown in (c). ABAB stacks, represented in blue, and ABCA, in black, are separated by the yellow lines which correspond to the domain walls. f) Close-up of the domain stacks: Each pair of single domain walls separating ABAB and ABCA regions merge into a double domain wall at the bottom, due to the elongation of the commensurate stacks.

# Results and discussion

The investigated sample, fabricated by dry transfer (see Methods section), consists of minimally twisted double bilayer graphene (TDBG) on top of a hexagonal boron nitride (hBN) flake on a silicon/silicon dioxide (Si/SiO$_2$) substrate. TDBG is composed of two AB-stacked (also called Bernal-stacked) bilayer graphene flakes rotated with respect to one another. The relative angle between the two graphene sheets is close to zero, in order to allow the formation of a moiré pattern with a large periodicity. A schematic of the final sample structure is shown in Figure 2a.

When stacking two graphene AB bilayers, three configurations are possible (see Figure 2b): ABAB (Bernal stacking), ABCA (rhombohedral) and ABBC. The latter is the most unstable[22] as the BB stacking (equivalent to AA) between the second and third layers is energetically unfavourable. Therefore, ABAB and ABCA regions tend to expand at the expense of ABBC. Thus, the resulting structure will consist of large commensurate ABAB and ABCA domains separated by transition regions, defined as saddle point (SP) boundaries.[23] This leads to the formation of a moiré pattern formed by discrete stacking domains (see Figure 2b). In addition, a small but finite energetic imbalance between the two commensurate domains exists: the Bernal stack is more stable than the rhombohedral one and therefore ABAB domains assume a convex shape.[23] More importantly, this intrinsic imbalance encourages the formation of double domain walls where two SP boundaries - single domain walls - merge into a unique domain wall separating two identical commensurate stacks.[13] This process is favored by the presence of (interlayer) strain, which in our case is introduced by the small twist angle.[13]

To first image the stacking domains, we use Piezoreponse Force Microscopy (PFM - see detailed description in the Methods section). The intralayer strain gradients introduced by twisting lead to an electromechanical coupling to the out-of-plane electric field and enable direct visualisation of the discrete stacking domains, as already shown in previous reports.[24,25] Figure 2c and d show the resulting PFM amplitude (c) and phase (d) images. A set of trian-gular discrete stacking domains can be observed. These are surrounded by higher contrast interfaces, corresponding to the SP boundaries. Convex and concave domains, ABAB and ABCA stacks respectively, can be also distinguished.[25] A schematic of the domain structure is depicted in Figure 2e and f, where ABAB (blue) and ABCA (black) stacks are separated by domain walls (yellow). From these images, we extract an effective twist angle between 0.07° and 0.15° using the relationship between the rotation angle $\vartheta$ and moiré wavelength $\lambda_m = (a/2) * \csc(\vartheta/2)$, where $a$ is the lattice constant of graphene.[26] We note that the moiré pattern in Figure 2c consists of stretched triangular domains. We attribute this observation to the combined action of two non-uniform strain components, respectively induced by atomic relaxation due to the small twist angle, and by external factors, such as wrinkles or the edge of the flakes.[12,13] Furthermore, it is possible to identify double domain walls separating two adjacent elongated ABAB stacks: these features originate from the union of two single domain walls surrounding the ABCA domain and merging into a unique boundary.[13] The contrast observed along the domain walls in response to the applied electric field in Figure 2c can be attributed to the flexoelectric component of the polarisation. Unlike piezoelectricity, flexoelectricity can be induced in centro-symmetric structures[27] as the non-uniform strain breaks the centro-symmetry. Such polarisation can originate from strain gradients induced by atomic relaxation and domain wall formation.[23,24]

UFM image of the region depicted in Figure 2c (see dashed cyan square) is shown in Figure 3a. Here, domain walls, corresponding to the incommensurate SP stacking, appear as dark lines. Almost no contrast variation is observed between neighboring ABAB and ABCA domains. We attribute the parallel lines in the lower half of Figure 3a to double domain walls. Each of these then bifurcates into two single domain walls with lower UFM signal strength that almost fades out at the apex of each stacking domain. The bifurcation point corresponds to the beginning of the ABCA domain (see Figure 2f). The measured width of the double domain walls is 12 nm, evaluated at full width at half maximum (FWHM).

By calibrating the UFM[17] (see Methods section and Supporting Infomation for details), we can extract absolute values of the local out-of-plane Young's modulus. We find a Young's modulus equal to $39 \pm 2$ GPa in the ABAB/ABCA stacking domains, while it decreases by $7 \pm 1$ GPa along the double domain walls (Figure 3b,c) and by $3 \pm 1$ GPa along the single domain walls. As discussed further, we attribute this softening to the lower interlayer coupling energy of the domain wall atoms coupled with their increased interatomic distance. This assumption is in good agreement with previous works on domain walls in ferroelastic and ferroelectric materials, which show a reduction of the saddle point energy in the domain wall, indicating that it is elastically softer.[28–30] It is worth mentioning that the absolute value of the softening along the double domain wall is larger than twice that of a single domain wall. This suggests an interaction between the domain wall and an atomistic reconfiguration whose description goes beyond the scope of this paper.

To explain the measured local variation in Young's modulus shown in Figure 3, force-field relaxation simulations are performed.[31–33] To this end, the Young's modulus is computed as the curvature of the interatomic potential energy with respect to the applied strain for the relaxed state (see Methods). The calculated bulk out-of-plane Young's modulus obtained for four layers of Bernal-stacked graphene is 35.5 GPa, which is close to the experimental value of 39 ± 2 GPa measured for graphite. In order to predict the local mechanical properties within the moiré superlattice of TDBG, an out-of-plane strain has been applied locally. The local out-of-plane Young's modulus was then estimated by scanning the calculations over the entire moiré cell. The simulated map of the Young's modulus for TDBG with a twist angle of $0.76°$ is presented in Figure 4a. Due to computation cost limit, smaller angles could not be explored and thus only single domain walls could be modelled as double domain walls appear at smaller twist angles.[13]

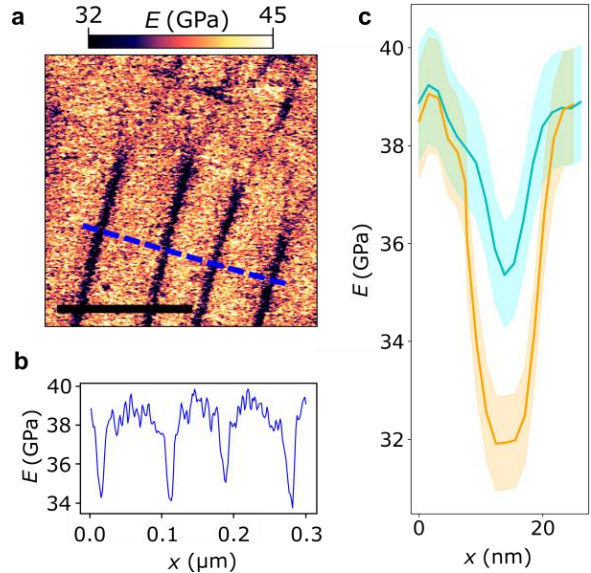

**Fig. 3. UFM visualisation of moiré pattern in TDBG.** a) UFM map of the TDBG moiré superlattice. Scale bar: 200 nm. b) Profile of the Young's modulus ($E$) (along the dashed line in a). c) Comparison between the profile of the Young's Modulus along a single domain wall (blue line) and double domain wall (orange line). Shaded areas correspond to the error on the line profiles. Errors were estimated by taking the minimum and maximum profile values.

The different stacking domains can be distinguished by their corresponding mechanical properties. Indeed, the ABAB and ABCA stacks are characterised by a higher stiffness than the SP and ABBC ones. The softening of the out-of-plane Young's modulus at the single domain walls is further illustrated in Figure 4b, corroborating the UFM measurements shown in Figure 3c (blue curve). Figure 4c illustrates the softening mechanisms occurring along domain walls. Due to the slight in-plane misalignment between the two bilayer graphene lattices along the domain walls (right panel in Figure 4c), vertical displacement is facilitated compared to the ordered, commensurate (ABAB) stacking order (left panel in Figure 4c). This is further illustrated by Figure 4d, which shows the interlayer coupling energy as a function of interlayer distance in ABAB stacks (red) and domain walls (blue). The difference between the two curves implies that there is a change in interlayer couplings when passing from a stacking configuration to another, leading to different response to vertical strain and a variation of Young's modulus.

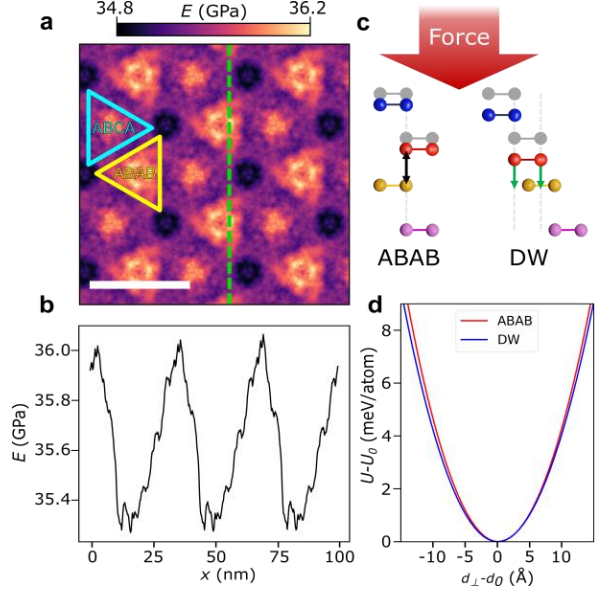

**Fig. 4. Theoretical Young's modulus for double bilayer graphene with a 0.76° twist angle.** a) Young's modulus $E$ map obtained after force-field relaxation modeling of the double bilayer graphene atomic structure. The different stacking orders are indicated by yellow (ABAB) and cyan (ABCA) triangles. Scale bar: 40 nm. b) Young's modulus line profile taken along the green dashed line illustrated in (a). c) Schematic illustrating the different displacement of the atoms in ABAB domains and domain walls under the action of an applied vertical force, i.e. the UFM tip. d) Interlayer coupling energy $U$-$U_0$ as a function of interlayer distance $d_\perp$-$d_0$ in ABAB stacks (red) and domain walls (blue).

# Conclusion

In this paper, we use two force microscopy techniques - PFM and UFM - to image the moiré pattern in minimally twisted double bilayer graphene and to investigate its mechanical properties. By means of PFM, we first observe the presence of non-uniform strain, manifesting as stretched triangular domains in the moiré pattern. We find evidence of strain-induced phenomena, in particular a flexoelectric effect along the domain walls and the generation of double domain walls. Then, by performing UFM measurements, we quantify the variation of Young's modulus along the various stacking domains, which results in a softening of $7 \pm 1$ GPa along the double domain walls. Such decrease in Young's modulus can be explained by force-field simulations which agree qualitatively with the experiments. Our findings confirm UFM as an effective technique to extract quantitative local mechanical information from twisted 2D materials. Such information will be valuable for improving device stability, investigating the impact of strain on optoelectronic properties or on the formation of exotic superconducting and magnetic phases.

# Methods

## Sample preparation

The twisted double bilayer (TDBG) stack was fabricated by the dry transfer technique with a Polycarbonate (PC) film (10% mass concentration) supported by a Polydimethylsiloxane (PDMS) bubble stamp. Two parts of a bilayer graphene, previously cut with an atomic force microscope,[34] were picked up successively with a relative twist angle of around 1°. The TDBG stack was deposited onto a hexagonal boron nitride flake (hBN) by transfer at room temperature without melting the PC film.[35] Such a cold transfer technique minimises the contamination in the surface exposed TDBG stacks. The TDBG stack was placed on the hBN at a sufficiently high angle to minimise the submoiré contribution that could arise from the hBN substrate.

## Scanning Probe Microscopy

All scanning probe microscopy measurements were performed on an AFM Multimode Nanoscope IV (Bruker). The lock-in detections were realised externally either with a Stanford Research System 830 or with a Zurich Instrument HF2LI lock-in amplifiers.

### Piezo Force Microscopy

To perform PFM measurements, an AC bias ($\sim$ 330 kHz and 1-3 V amplitude) is applied between the conductive AFM tip and the sample to produce an electric field along the vertical direction. The frequency is set to the contact resonance frequency between the tip and the sample. This induces a periodic piezoelectric strain on the sample that is detected by measuring the torsion (lateral-PFM) or deflection (vertical-PFM) of the AFM lever at the frequency of the contact resonance of the AFM tip. Here, we used vertical PFM measurement to image the moiré pattern in TDBG.

**Ultrasonic Force Microscopy**

To perform UFM measurements, the sample is mechanically bonded to a piezoelectric transducer using a crystalline compound, phenyl salicylate, maximising the ultrasonic contact.[17] The tip used is a regular contact mode silicon tip (Budget Sensors ContAl-G, k = 0.2 Nm$^{-1}$, $f_0$ = 13 kHz). By applying a modulated AC bias ($f \approx$ 4 MHz, modulated at ~ 2.7 kHz) to the piezoelectric transducer, the sample is vibrated vertically (see Figure 2a). Since the excitation frequency of the sample is much higher than the fundamental resonance frequency of the cantilever (typically 3–10 kHz), it is unable to follow the sample vibration. The cantilever therefore becomes inertially stiff.[16] For this reason, when in contact with the surface, the tip can indent into the sample even when the latter is much more rigid than the cantilever, making this technique extremely sensitive to the variations of the elastic properties of the sample.[36] The UFM response is obtained by modulating the amplitude of the ultrasonic vibration at 2.7 kHz, which is a frequency value lower than the first cantilever resonance and higher than the feedback cut-off (around 1 kHz). Being the vibration amplitude cyclically varied by the modulation, the system oscillates between two regimes, linear and non-linear respectively, separated by the critical value of the ultrasonic amplitude at which the contact between the tip and the surface breaks. When entering the non-linear regime, a rectification effect that physically consists in the additional positive deflection described above is produced on the cantilever.[18] To quantify the local mechanical properties, the UFM response was calibrated by comparing measured signals on reference materials with known Young's modulus: Silicon dioxide ($SiO_2$), hexagonal boron nitride (hBN) and graphene (see Supplementary Information). This allows us to convert the changes in UFM signal to variations in Young's modulus. While UFM enables the study of surface and subsurface material prop-erties with nanometric resolution,[17,37] it has to be taken into account that such response is predominantly sensitive to elastic (modulus) and adhesive properties.[19] To rule out any adhesion effects, we performed adhesion maps of the same region using PeakForce Quantitative Nanomechanics (QNM) tapping. No variation of the adhesion can be observed (see Supplementary information).

**Force-Field relaxations simulations**

In this work, the atomic structure relaxation was computed using a classical force-fields model.[38] In particular, intralayer forces are determined by the optimised Tersoff and Brenner potentials,[39] while interlayer forces are modelled using the Kolmogorov–Crespi potentials.[31,33] In addition, in order to mimic the experimental setup consisting of twisted double bilayer graphene on top of a hexagonal boron nitride flake (i.e., a flat substrate), the bottom graphene layer was kept flat in our relaxation simulations. The atomic structure was then optimised until force components are smaller than 0.5 meV/atom. The mechanical properties of twisted double bilayer graphene were considered for the obtained relaxed lattice. In particular, the Young's modulus is computed as the curvature of the curve of interatomic potential energy $U$ with respect to the applied strain $\epsilon$ at the relaxed state: $E_Y = V^{-1} \partial^2 U / \partial \epsilon^2$ where $V$ is the system volume. To model the local mechanical properties of the considered moiré superlattice, we then locally applied an out-of-plane strain at the point $\mathbf{r}_0$ using a Gaussian-like expression $\epsilon(\mathbf{r}, \mathbf{r}_0) = \epsilon_0 \exp(-|\mathbf{r} - \mathbf{r}_0|^2 / 2\sigma^2)$. The local out-of-plane Young's modulus was then investigated by scanning the calculation, i.e. displacing the computed point $\mathbf{r}_0$) over the cell.

# Author Contributions

J.S. and P.G. conceived and supervised the experiments. A.C., supported by S.G.-M., K.A. and O.K., performed and evaluated the UFM measurements. P.d.C., B.N. and B.H. per-formed the PFM measurements. Y.H., S.M. and R.R.-P. fabricated the samples. K.W. and T.T. provided the hBN. V.-H.N. and J.-C.C. performed the force-field relaxation calculations. The manuscript was written through contributions of all authors. All authors have given approval to the final version of the manuscript.

# Conflicts of interest

There are no conflicts to declare.

# Acknowledgements

We acknowledge financial support from the F.R.S.-FNRS of Belgium (FNRS-CQ-1.C044.21-SMARD, FNRS-CDR-J.0068.21-SMARD, FNRS-MIS-F.4523.22-TopoBrain, FNRS-CR-1.B.463.22-MouleFrits, FNRS-PDR-T.0029.22-Moiré), from the Federation Wallonie-Bruxelles through the ARC Grant No. 21/26-116 and from the EU (ERC-StG-10104144-MOUNTAIN). This project (40007563-CONNECT) has received funding from the FWO and F.R.S.-FNRS under the Excellence of Science (EOS) programme. This work was also partly supported by the FLAG-ERA Grant TATTOOS, through F.R.S.-FNRS PINT-MULTI Grant No. R 8010.19. We acknowledge the technical support from Bruker UK. The support of the European Union's Horizon 2020 Research Project and Innovation Program–Graphene Flagship Core3 (No. 881603), EPSRC EP/V00767X/1 HiWiN project is fully appreciated. Computational resources have been provided by the CISM supercomputing facilities of UCLouvain and the CÉCI consortium funded by F.R.S.-FNRS of Belgium (No. 2.5020.11). R.R.-P. acknowledge the ERC starting grant TWISTRONICS. This work was supported by the French RENATECH network and the DIM-SIRTEC. K.W. and T.T. acknowledge support from JSPS KAKENHI (Grant Numbers 19H05790, 20H00354 and 21H05233).